# A Quality-of-Service Compliance System using Federated Learning and Optimistic Rollups

João Paulo de Brito Gonçalves
Federal Institute of Espirito Santo (Ifes)
Cachoeiro de Itapemirim - ES, Brazil
jpaulo@ifes.edu.br

Guilherme Emerick Sathler
FACIG
Manhuaçu – MG, Brazil,
emericksathler@gmail.com

Rodolfo da Silva Villaca
Federal University of Espirito Santo (Ufes)
Vitoria - ES, Brazil

**Abstract.** Edge computing brings a new paradigm in which the sharing of computing, storage, and bandwidth resources as close as possible to the mobile devices or sensors generating a large amount of data. A parallel trend is the rise of phones and tablets as primary computing devices for many people. The powerful sensors present on these devices combined with the fact that they are mobile, mean they have access to data of an unprecedentedly diverse and private nature. Models learned on such data hold the promise of greatly improving usability by powering more intelligent applications, but the sensitive nature of the data means there are risks and responsibilities to storing it in a centralized location. To address the data privacy required for some data in these devices we propose the use of Federated Learning (FL) so that specific data about services performed by clients do not leave the source machines. Instead of sharing data, users collaboratively train a model by only sending weight updates to a server. However, the naive use of FL in those scenarios exposes it to a risk of corruption, whether intentional or not, during the training phase. To improve the security of the FL structure, we propose a decentralized Blockchain-based FL in an edge computing scenario. We also apply blockchain to create a reward mechanism in FL to enable incentive strategy for trainers.

## 1 Introduction

Federated Learning (FL) is an emerging distributed learning framework that allows multiple devices to distributively train a shared model without the need of collecting and aggregating all data from all the nodes in the network, thereby protecting the data privacy and reducing the communication overhead. FL preserves user privacy and data security by design since no data transfer is required, what it is interesting for network operators who have in their possession sensitive data about organizations and users that don't want to share or publish data from their edge devices. FL also enables several parties to collaboratively train a ML model so that each of the parties can enjoy a better model than what it can achieve alone.

However, there are several main concerns about the current FL systems:

1. due to the distributed nature of these systems there is a need for coordination in the learning process generally by a central server. However, this central server can easily become a single point of failure and may not be trustworthy;

2. in the first FL algorithm proposed [30], trainers collaborate voluntarily in the system, which may be feasible in a network controlled by the system administrator but in a context where it is necessary for devices to volunteer to participate in the system, such as the Internet, some reward mechanism is needed to encourage participants to enter the system;

3. a very common and challenging kind of attack FL systems is when an attacker follows the system protocol but propagates arbitrary malicious information to benign system participants, aiming to degrade the system performance and further mislead or control the system output [50].

Thus, to its decentralized nature, the blockchain technology can increase the FL model security and performance. It allows all members of the system to be aware of the current (and past) network resource availability. A secure resource exchange is guaranteed by smart contracts and distributed consensus algorithms, allowing the system to evolve autonomously without the need of centralized authorities [11]. The greatest benefits that would be achieved by applying blockchain in FL are [47]:

– the immediate updates made by each participant to its local model can be chained together on the distributed ledger offered by a blockchain such that those model updates are audited.
– model updates can be chained in a cryptography way such that their integrity and confidentiality can be guaranteed;
– the tokenization structure offered by some blockchain platforms allows the creation of incentive mechanisms inherent in the structure itself, which allows automating reward mechanisms;
– every model update, be it either local weights or gradients, can be traced to and associated with an individual participant, which helps the detection of tamper attempts and malicious model substitutes.

Service Level Agreements (SLA) are formal contracts between consumers and service providers that outline the parameters of some IT services to be supplied [46]. Besides the nature of the service itself and its expected performance level, a SLA also specifies the procedures for monitoring and reporting problems, time limits for problem resolution and the consequences for both clients and service providers when clauses are violated. In the network domain, a SLA can be established between consumers and network operator, or between two operators. Network SLAs are typically conceived in terms of network performance between exchange points where the service is provided. They can cover physical network aspects, such as bandwidth, availability, latency, jitter, packet loss or error rate [46]. However, there are SLAs that are not related only to network performance requirements, but also to different types of services in clouds (such as computing units, data storage, hosting, etc). If one of these requirements is violated (i.e., not met or delivered), the provider has to compensate the customer according to the agreed terms, which are also defined in the SLA.

Monitoring and reporting the QoS allows clients and service providers to check how well is the service complying with the SLA. If the specified performance parameters are violated during operation of the system, the SLA constitutes a legal foundation that eventually serves as a means of dispute resolution between service providers and customers. As examples of violations, we can cite: unannounced connection interruptions, connection bandwidth pro- vided far below the minimum agreed, unavailability of a service, access latency above the agreed, etc.

As a validation case for the proposal, we will use the compliance QoS verification case. Thus, this work has as main objective to propose and develop a framework to deploy a blockchain empowered FL model to verify compliance with QoS parameters. We simulated a edge computer architecture using the Flower framework and interacting with the Optimistic Rollup implementation from Cartesi Network in order to apply blockchain technology in a scalable way.

The rest of this paper is organized as follows: in Section 2, the background about the main technologies used in our paper is presented. Section 3 provides the related work and discuss the proposal regarding the state-of-the-art of the use of blockchain and FL. In Section 4 we present our proposal while Section 5 presents the implementation and preliminary results. Finally, in Section 6, we conclude the paper and discuss future work proposals.

## 2 Background

### 2.1 Blockchain

Despite its initial financial original application [32], blockchain technology has grown to a multiplicity of different applications, where there is the constant need for unchanging and distributed recording of operations performed on a system. The blockchain is a distributed ledger, where each participant has a copy of the database with all the validated information. Besides, a consensus protocol is implemented among the participants in order to allow them to agree about the global state of the blockchain.

In a blockchain, each block is a set of transactions chained through hash addresses. Each block includes, among other information, a timestamp, its hash, and the hash of the previous block, so that once the block is created, it cannot be tampered under the penalty of the stored hash not matching to the hash of the modified block, thus evidencing the attempted fraud. This tampering becomes more difficult the longer is the chain of blocks due to the computational power required for the task.

In order to confirm a transaction presence in a given block, a data structure called Merkle Tree [31] is generally used. In this structure new transaction data is stored with pointers to the original block locations for unchanged data. Trans- actions are repeatedly paired, merged, hashed, and remade until only one hash (the Merkle root), remains. Each subsequent block saves the Merkle root of the previous block.

In public blockchains, i.e., where access is not controlled by a central authority, validation of transactions and blocks is often based on a consensus protocol. The initial consensus protocol proposed by Nakamoto was Proof of Work (PoW). In PoW, a cryptic challenge is proposed in order to create a valid block, once solved, the block is propagated over the network. Only after a transaction has been validated (included in a valid block), it is performed, which might change the blockchain state.

PoW was proposed to discourage malicious users from creating fraudulent transactions on the network, but there is criticism regarding performance loss caused by their use. In the past few years, other validation strategies has been proposed, as Proof of Stake [24] and Proof of Authority [37] among others, based more in currency stake and reputation respectively and less in computation power. At this point, it is important to differ between two basic types of blockchains: public and permissioned.

In public blockchains, any computer can join the network and have full access to it. Because of this anonymous character of computers, measures to mitigate attacks must be adopted which results in performance degradation, as the Proof of Work by example. Bitcoin and Ethereum [49] are examples of public blockchains. Permissioned blockchains are mainly used in corporate environments. This means that a user must have a certain level of access to interact on this network, read transactions, and participate in the consensus process. Hyperledger Fabric [14] is an example of a permissioned blockchain.

## 2.2 Ethereum and Smart Contracts

Several blockchain platforms have emerged in recent years and among the most popular are those based on the Ethereum platform. Ethereum is a platform for executing blockchain applications that are modeled as smart contracts and has its own cryptocurrency, the ether.

Smart contracts capture and translate traditional legal contract clauses into a series of computational rules which are executed automatically and, once validated, don't require additional legal instruments [41].

A fee is paid for transactions to be included by miners for mining. If this fee is too low, the transaction may never be picked up; the more the fee, the higher are the chances that the transactions will be picked up by the miners for inclusion in the block. Conversely, if the transaction that has an appropriate fee paid is included in the block by miners but has too many complex operations to perform, it can result in an out-of-gas exception if the gas cost is not enough. In this case, the transaction will fail but will still be made part of the block and the transaction originator will not get any refund.

On the Ethereum platform, applications run on the Ethereum Virtual Machine (EVM), which executes smart contract instructions, allowing you to enter and query stored data. It is completely isolated, and the code that runs on it has no access to any external resources such as the network or the computer file system [15]. Ethereum now uses Proof of Stake as its current consensus mechanism after its transition phase towards Ethereum 2 as well as the use of Sharding (each node having only a part of the data on the blockchain, and not all the information). These two main changes should enable the processing of up to 10000 transactions per second [40].

The Ethereum platform is currently the largest general purpose public blockchain exponent on the Internet. It is a very flexible alternative to the development of dApps (Decentralized Applications) as it provides a complete programming language, different from the Bitcoin platform, which has a very limited scripting language and is used just to support basic and necessary network operations [15]. A dApp is a decentralized application that uses a smart contract in the blockchain as backend, and a web interface as frontend, allowing users to insert and receive data from the blockchain in a friendly way.

## 2.3 Oracles

Oracles are an important component of the smart contract ecosystem. The limitation with smart contracts is that they cannot access external data which might be required to control the execution of the business logic [5]. Usually smart contracts need information that is processed outside their computational logic and the availability of this information is crucial to the achievement of smart contracts full potential. However, this is challenging since smart contracts can only access and write information that is stored on the blockchain, which is an enclosed network without direct interfaces to the real world. Oracles bridge the gap between the blockchain and the real-world by feeding data from outside the blockchain to smart contracts. They are usually application's APIs which produce data that can be consumed by smart contracts. They are used to report events and data created after the smart contract has been programmed [10].

There are two types of oracles: centralized and decentralized. The centralized oracles suffer from the same problem of centralized networks, they have a single point of failure, besides that capture data directly from third-party service providers, websites, or different types of API and then report the results to smart contracts; this makes it possible to manipulate the results. Hyperledger Avalon [23] deviates intensive processing from the main blockchain to an off-chain channel to support centralized but trustable oracles (by using trusted execution environments). Since multiple blockchains can use the same data, it stimulates interoperability.

The decentralized oracles are designed so that they do not suffer from a single point of failure problem [48]. One of the most efficient ways to achieve this is to build multiple oracle servers and make them continuously running even when an oracle server fails. Examples of decentralized oracles are: Chainlink [17], Witnet [12] and Dos Network [33].

## 2.4 Blockchain Interoperability and Layer2-Solutions

Interoperable blockchains open up a world where moving assets from one platform to another, or payment-versus-payment and payment-versus-delivery schemes or accessing information from one chain inside another becomes easy and even implementable by third parties without any additional effort required from the operators of the base blockchain protocols.

One important application of blockchain interoperability is the scalability problem. Using two or more integrated blockchains, we can split the processing using a network with higher throughput capacity for the most onerous tasks and using the public blockchain for storage and simpler tasks. We call these solutions Off-Chain Protocols [16].

In general, off-chain protocols belong to one of the following four categories, described below.

- Channels: are fully secured by a blockchain system such as Ethereum and work only for a specific set of applications.
- Commit-chains: rely on one central intermediary who is trusted regarding availability.
- Sidechains: are usually EVM-compatible and can scale general-purpose applications but are less secure because they do not rely on the security of the underlying blockchain and thus create their own consensus models.
- Rollups: solutions that push complex computations off-chain, meaning that they run on a separate computing environment (layer-2) outside of the main network (layer-1, such as the Ethereum network). There are two main types of rollups: ZK-Rollups (Zero Knowledge-Rollups) [9] and Optimistic Rollups [43]. In ZK-Rollups, transactions are processed off-chain and then added to Ethereum as a single, compressed transaction. The security of the off-chain computations is ensured using zero-knowledge proofs. In Optimistic Rollups transactions are batched and processed off-chain and then added to Ethereum unaccompanied by proofs, they're proposed and, if not challenged, confirmed on-chain. Challenging a state update proposal is done using fraud proofs.

### 2.4.1 Optimistic Rollups

Optimistic rollup is a layer-2 scaling solution for Ethereum, where transactions are processed off-chain by a set of validators and then period- ically merged on-chain. This allows for faster and cheaper transactions compared to executing them directly on Ethereum's network while maintaining the security of the underlying blockchain. The optimistic aspect refers to the assumption that the validators will act honestly and that any malicious behavior can be detected and challenged on-chain [43].

By using Optimistic Rollups technology, we can use the advantages of blockchain technology without the drawbacks of public blockchains. As main advantages we can cite:

- Scalability: Optimistic Rollup can process a high number of transactions per second.

- Cost-effective: Transactions on Optimistic Rollup are cheaper than on Ethereum's main chain, as they are processed on a layer-2 solution.
- Compatibility: Optimistic Rollup is fully compatible with Ethereum, meaning that it supports the same programming languages and tools.
- Security: Optimistic Rollup uses cryptographic techniques to ensure that transactions are processed correctly and securely.
- Speed: Transactions on Optimistic Rollup are processed much faster than on Ethereum's main chain, due to the use of layer-2 solutions.
- Interoperability: Optimistic Rollup enables interoperability between different Ethereum-compatible networks and applications.

Optimistic rollups rely on a fraud-proving scheme to detect cases where transactions are not calculated correctly. After a rollup batch is submitted on Ethereum, there's a time window (called a challenge period) during which anyone can challenge the results of a rollup transaction by computing a fraud proof.

If the fraud proof succeeds, the rollup protocol re-executes the transaction(s) and updates the rollup's state accordingly. The other effect of a successful fraud proof is that the sequencer responsible for including the incorrectly executed transaction in a block receives a penalty.

### 2.4.2 Cartesi Rollups

Cartesi Rollups are a layer-2 scaling solution for Ethereum, designed to increase the scalability and lower the cost of decentralized applications using Optimistic Rollups. They work by executing transactions off-chain, on a virtual machine with a Linux-based operating system that is EVM-compatible. The result is then verified and committed on-chain. This allows for faster and more affordable transactions, while still maintaining the security guarantees of the underlying Ethereum network. Cartesi Rollups can also handle complex computation and data storage, making them suitable for a wide range of decentralized applications. It allows DApps to be developed using conventional programming languages by offering a Linux operating system coupled with a Blockchain infrastructure. Smart contracts for Ethereum must be written using Solidity but using Cartesi, we can write it with more familiar languages such as Python. Figure 1 shows the architecture for a Dapp in Cartesi [42].

### 2.5   Federated Learning

As a novel distributed Machine Learning approach, Federated Learning (FL) is a promising technology. In FL, each distributed client trains the local model on the local database, and then sends the local model gradient to the central coordinator for global model optimization. Compared with traditional FL, in FL, the client only needs to send the local model gradient without revealing the local raw data to any other participants. In a nutshell, the parts of the FL algorithms that touch the data are moved to the trainers' computers. FL is not just limited to Supervised Learning. It can also be used in other areas, such as Unsupervised Learning [28].

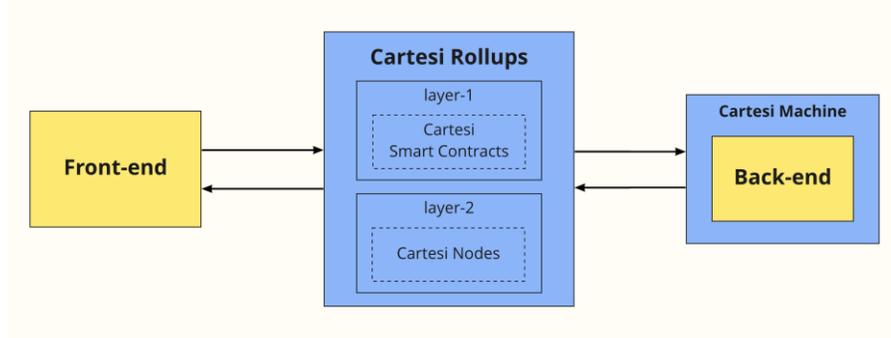

**Fig. 1.** Cartesi Architecture

The concept of Federated Learning was first proposed by McMahan et al. [30] who developed the FedAvg algorithm [29] and was first applied in [51]. This concept enables edge computing from the machine learning perspective. In a border definition of FL, multiple clients are coordinated to train model locally and aggregate model updates to achieve the learning objective, while raw data are stored in clients and not exchanged or transferred. These updates are much harder to interpret than pure data, so this is a major improvement for privacy. For some applications with huge amounts of data, it might also be cheaper to communicate the updates compared to directly sending the data. Because of this new feature, it has been seen as a very promising technique to be applied in sensitive data security, where data privacy is essential [1], [45]. In a more general sense, FL allows machine learning algorithms to gain experience from a broad range of data sets located at different locations. The approach enables multiple organizations to collaborate on the development of models, but without needing to directly share secure data with each other.

Compared with conventional ML algorithms, FL enhances the user privacy data by executing the training process locally and improves the performance of training results by involving many users to train a global model collaboratively.

In FL, the gradients represent the evolution of the model weights after a training step. Consider Wi a specific weight of a model and Gi a specific gradient achieved because of training. Usually, a training step involves the following steps:

- fetch the current version of the weights Wi and apply it to the known architecture.
- trains the model locally with its own dataset, generating new gradients Gi + 1.
- uploads the resulting gradients Gi + 1.

This way, the dataset used for training never leaves the partner's infrastructure, ensuring its privacy and excluding any leakage of the processed data. Because access to the previous datasets is forbidden, this training step builds epochs only from the new dataset provided by the user. After this training, the generated gradients Gi + 1 are uploaded and applied to the previous weights (Wi), leading to new parameters (Wi + 1) and a candidate model, noted tmp(Mi + 1). This structure can be visualized in Fig 2 where the coordinator is the aggregation server which sends an initial model to the local data owners (a.k.a trainers). Each trainer trains a model using their respective dataset and send the model weight updates to the aggregation server. The aggregation sever then combines the model updates received from the trainers.

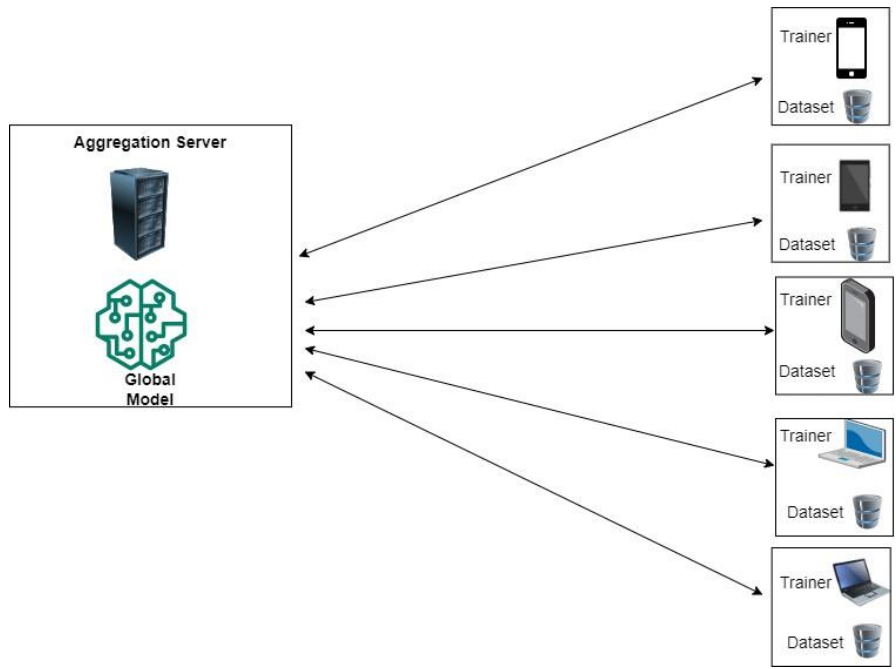

**Fig. 2.** Operational Structure of Federated Learning

To protect against any leaks coming from the gradients, it is possible to share a small proportion of the gradients and add some noise as suggested by [39] in order to improve differential privacy. The parameter E, controlling the trade-off between the accuracy and privacy, should be determined by the user.

### 2.5.1. Federated Averaging (FedAvg)

The Federated Averaging (FedAvg) algorithm was first employed for FL over horizontally partitioned datasets, in the context of updating language models on mobile phones [29].

In FedAvg, each party uploads clear-text gradient to a coordinator (or a trusted dealer, or a parameter server) independently, then the coordinator computes the average of the gradients and update the model. Finally, the coordinator sends the clear-text updated model back to each party. The central server repeats these procedures many times for improving the global model. This cyclic training process continues until the model accuracy reaches a saturation point [30].

Although FedAvg has shown great advantages in enabling collaborative learning while protecting data privacy, it still faces an open challenge of incentivize people to join the federated learning system by contributing their computation power and data. Besides that, FedAvg is not suitable for industrial IoT systems since the heterogeneity between the different client local data sets exists heavily in industrial IoT systems (i.e., the distribution and size of the data set on each client may be different) [52]. Take the air conditioner failure detection as an ex- ample, the deployed air conditioners might have different working environments (e.g., weather) and different users.

Fig. 3 shows a basic interaction in FedAvg. At the beginning of an iteration, a subset of K clients are randomly selected by the server. They receive a copy of the current model parameters and use their locally available training data to compute an update. The update of the i-th client is denoted by Hi. The updates are then sent back to the server.

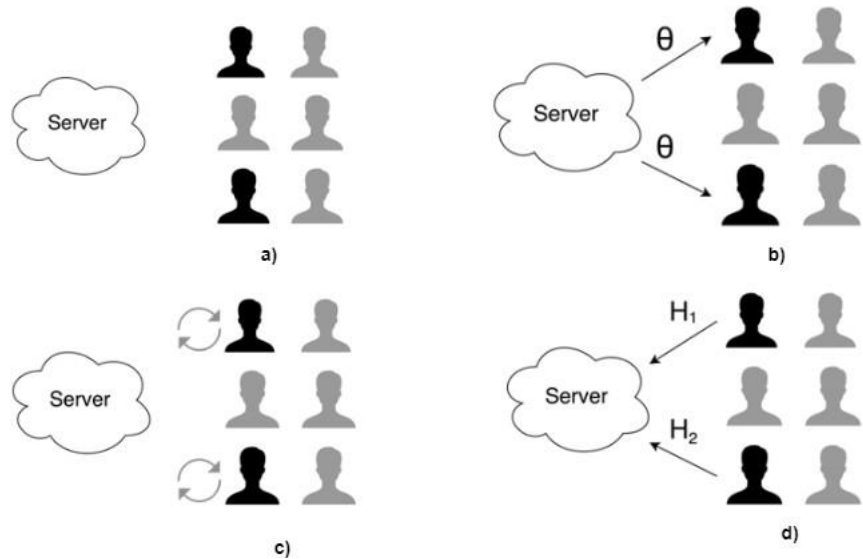

**Fig. 3.** One communication round in a Federated Learning system with (a) the server selects K users; (b) they receive the current model; (c) and compute updates using their data (d) updates are shared with the server

Surprisingly, a huge part of machine learning is based on a conceptually sim- ple method called Stochastic Gradient Descent (SGD). It is an iterative algo- rithm where the model parameters are repeatedly moved into a direction where they work a little bit better. In the beginning, the model is initialized with an arbitrary set of parameters.

Afterwards, the iterative optimization process begins. In each iteration, the partial derivatives of the loss with respect to the model parameters are computed. To be able to do this, SGD requires the prediction function f and the loss function to be differentiable.

After each round, the quality of their proposed updates should be improved as much as possible. Generally, this works by increasing the amount of work performed on each user's computer.

In the case of gradient descent, this idea can be implemented by performing multiple update steps locally before sending anything to the server. By running several iterations, the update that is computed can be of much higher quality. This is often the case because it adds degrees of freedom to the update. only taking one step, the update must be alongside the gradient of that iteration. By allowing several iterations, the update can move in different ways and has more chances of properly going into an optimum.

The server waits until it has received all updates and then combines them into one final update. This is usually done by computing an average of all updates, weighted by how many training examples the respective clients used.

## 3   Related Work

Po-FL [35] is a novel energy-recycling consensus algorithm where the cryptographic puzzle in PoW is replaced with federated learning tasks. To realize PoFL, a general framework is introduced, and a new PoFL block structure is designed for supporting block verification. To guarantee the privacy of training data, a reverse game-based data trading mechanism is proposed, which takes advantage of market power to make a rational pool maximize his utility only when he trains the model without any data leakage, thus further motivating pools to behave well.

Qu et al. [36] propose a novel blockchain-enabled FL (FL-Block) scheme to be used in a Fog Computing scenario. FL-Block allows local learning updates of end devices exchanges with a blockchain-based global learning model, which is verified by miners.

Li et al. [27] combine blockchain with FL and propose a crowdsourcing framework named CrowdSFL, that users can implement crowdsourcing with less overhead and higher security, addressing the problem of security issues in crowd computing.

The Cosmos network [25] is a decentralized network of independent parallel blockchains, called zones. Zones can transfer data to other zones directly or via hubs, which minimize the number of connections between zones and avoid double spendings. For integration between the Cosmos Hub and the Ethereum blockchain, the Gravity bridge is already under development.

Unlike Cosmos, which has its own blockchain network, Polygon [8] uses Ethereum to host and execute any mission-critical component of their logic.

It has as proposal to use Ethereum security features and to improve Ethereum limitations as low throughput. Polygon provides interoperability to exchange arbitrary messages with Ethereum and other blockchain networks. Currently, Polygon is developing a suite of zero-knowledge rollups (ZK rollups) to increase throughput on Ethereum without sacrificing decentralization or security. An example is Polygon Zero which adopts a different approach by generating proofs simultaneously for every transaction in the batch. Then the machine aggregates multiple transaction proofs into a single proof submitted on the Ethereum network. Polygon Zero is designed to be compatible with the Ethereum Virtual Machine (EVM) and can batch up to 3,000 transactions per block [9].

Another example of Blockchain Connector is Hyperledger Cactus [6]. Cactus operates for each connected blockchain a group of validator nodes, which as a group provides the proofs of the state of the connected ledger. The group of validator nodes runs a consensus algorithm to agree on the state of the underlying blockchain, since a proof of the state of the blockchain is produced and signed with respect to the rules of the consensus algorithm. The framework provides examples of intercommunication between Hyperledger Fabric, Sawtooth [38] and Ethereum blockchains.

## 4    Proposal

Our framework proposal was designed with three layers: Training, Validation and Management. All the layers are connected to the blockchain, so that nodes can participate in reward mechanisms. This model is presented in Figure 4.

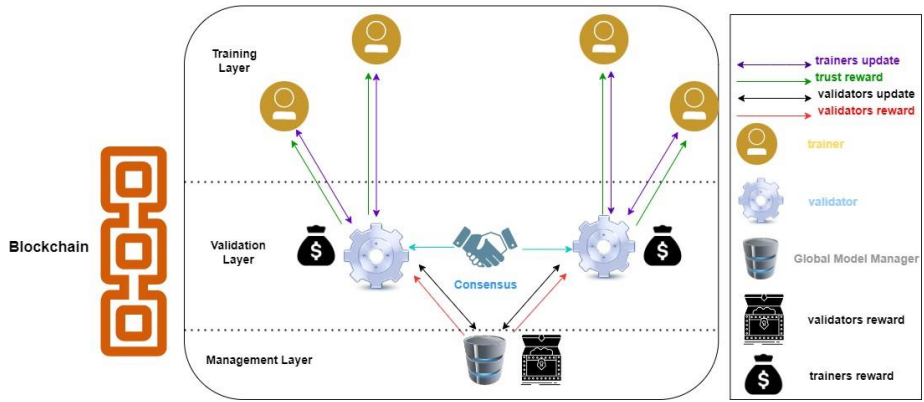

**Fig. 4.** Blockchain Federated Learning Structure

The first layer, Training, is made of the devices that perform the training with the local data and send their gradients $\mathbf{Gi + 1}$ to their corresponding validator. The second layer, Validation, is made of servers that validate the contributions of trainers by marking them as Positive and rewarding them if they are valid. Besides, the validators will calculate their own model **tmp(Mi + 1)**. using the valid contributions of the trainers and reach a result by calculating the average of the results obtained by the validators.

The third and last layer, Management, is composed by the Global Model Manager, which is the server responsible for distributing the model to the validators, sending the rewards that will be paid to the trainers and distributing the reward to the validator that makes the most accurate contribution, that is, closer to the average obtained.

The process begins when the validators send an initial model **Gi** as well as a public key to the miners attached to him. Each trainer individually calculates the gradient value based on the private training data and other hyper parameters received from the validator, which is encrypted using the received public key. Once a local model update is received by a validator **Gi + 1**, it decrypts the contribution and runs training using its own local dataset for a single epoch to evaluate the accuracy of the received model. If the model differs at most by 25% from the model trained by the validator itself, so the validator will mark this contribution as Positive and will send a reward to the trainer and the new model. If the model differs by more than 25% from the model trained by the validator itself, then the contribution will be discarded. We call this predefined threshold as Maximum Acceptable Deviation (MAD) which defines the baseline of distinguishing malicious from legitimate contributions. The algorithm to classify each contribution is described in Algorithm 1.

After that, the contributions marked as positives by each validators will be broadcasted for all validators in order to compute the models based on these contributions **tmp(Mi + 1)**, aggregating them for establishing a shared quality-improved model. After that, the average of the contributions of each validator will be calculated to reach the result.

---

**Algorithm 1** Rewards Analysis

---

**Input** : Contributions c, Model M
**Output:** Classification
**Result:** Contribution Classification

1 **for**(each round r) {
2 **for**(each trainer t) {
3 **for**(each contribution c received) {
4 mad ← recalcaccuracy(M);
5 **if** $c \geq mad$ **then**
6 | Classification ← Positive;
7 | t ← reward;
8 **else**
9 | Classification ← Negative;
10 **end**
11 return Classification   } } }

---

## 4.1   Incentive Mechanism

In federated learning, motivating data owners to continue participating in a data federation is an important challenge. The key to achieving this objective is to device an incentive scheme that shares the profit generated by a federation with participants in a fair and just manner. Before this step can be achieved, a mechanism for evaluating the contribution toward the federated model by a given data owner must be established.

Thus, the incentive mechanism plays a crucial role in ensuring that the trainers conduct their actions properly. To this end, the incentive mechanism needs to incentive participants via a reward scheme and penalize malicious behavior.

Without a proper incentive mechanism, clients may not be willing to join the data training which would reduce the scalability of the designed FL system [4].

In our proposal there are two types of rewards: the first one is used by validators to pay for contributions marked as *Positive* and the latest is used to reward the validator that had the best contribution to the model and will be used to choose who will mine the block. The stake of each device (trainers and validators) is recorded in the blockchain and recognized by the devices in the network because is inserted in the next block. Because of the validation mechanism, malicious devices are less likely to frequently receive the rewards for malicious contributions nor they will influence the training of the global model.

The number of tokens provided by the Global Model Manager for payment of trainers is fixed for each round and is decremented from the amounts spent to pay for valid contributions. The remaining amount value is recorded on the blockchain for later audits. The Manager uses the blockchain both to send tokens to validators to reward trainers and to reward the validators themselves and to save the final accuracy of the model at the end of the computation. The Manager can also, at any time, consult the stakes of the validators to select which node will mine the next block on the blockchain and the last results.

Unfortunately, there are two main difficulties that make traditional incentive mechanisms unfit in federated learning. First, computing nodes do not share their decisions due to privacy concerns. Without the information of other nodes, it would be impossible for a participant to derive an optimal decision with closed-form expression. Second, it is difficult to evaluate the contribution of participants to the accuracy of trained models. Evidence have shown that the relationship between model accuracy and the amount of training data is nonlinear [34]. The model accuracy depends on the model complexity and data quality and can hardly be predicted in advance Without accurate evaluation of contributions, incentive mechanisms cannot correctly reward participants, leading to financial loss or low participation rate.

## 4.2   A New Consensus Mechanism for Blockchained Federated Learning Systems

Although PoW is currently the most widely used consensus algorithm in blockchain platforms and its reliability has been extensively verified, PoW is not without its flaws. For instance, its large energy consumption has been criticized, and the centralization caused by mining pools has also been controversial. To address these problems, the PoS consensus mechanism were proposed. In the PoS mechanism, each entity use coinage as a measure of its equity. The coinage is defined as coin × t, where t represents time. The more stake an entity has, the more likely it is to become the next block producer. As a result, the PoS no longer requires entities to perform many hash operations, which greatly reduces the energy consumption [44].

It is important to highlight that to maintain the blockchain's security, a consensus mechanism must satisfy the following properties [19]:

– Persistence: Once a transaction is confirmed by an honest user, all other honest users will also confirm that transaction, and the transaction's position in the blockchain is the same for all honest users.
– Liveness: After a sufficient period, a valid transaction will be confirmed by all the honest users.

Our proposal is inspired by PoS [24], which allows the validator that has made the most accurate contributions among the validators at end of FL to receive reward proportional to the completion of this task, increasing its stake. Our system comprises two types of nodes: trainer and validator. Validators are responsible for maintaining models on the underlying blockchain network through decentralized consensus.

In our proposal, the validator with the greatest stake will be chosen to mine the next block that will be inserted into the blockchain. In the case of a stakes tie between two or more validators, the tiebreaker is given by the validator who has made the best contribution in the current round. The criteria for define which is the best contribution is to choose the contribution closest to the average obtained by the calculus of the contributions of all validators. The validator author of this contribution will win the reward in that round. This steak increase based in good contributions is similar to the concept of ranking or reputation [21].

We call our consensus mechanism Federated Learning Proof of Stake (FLPoS) and it be implemented using the rewards mechanisms from the Cartesi Network in the next phase of the work.

## 5    Proposal Evaluation

In order to show that the architecture is resilient to malicious users, we conducted comparative simulations in scenarios using the proposed architecture and a federated learning system without the use of additional security mechanisms.

Specifically, we have used a Stochastic Gradient Descent (SGD) [2] algorithm to learn a general predictive model.

For FL implementation we have used the Flower Python's framework [7]. Flower supports heterogeneous environments including mobile and edge devices, and scales to many distributed clients. Trainers are allowed to train for as long as a round may extend, therewith, until a new model is released, trainers can keep advancing local training, periodically sharing it with a validator. This framework uses the centralized architecture for federated learning, that is, a server does the client orchestration and model aggregation. In each round the calculated weights are recorded in .npz files. In this phase we are still using a generic dataset for ML systems [13], but in the next phase we will use a dataset collected from a Internet Service Provider (ISP) using a distributed Oracle and we want to predict crowds on 5G cells and variations in the QoS.

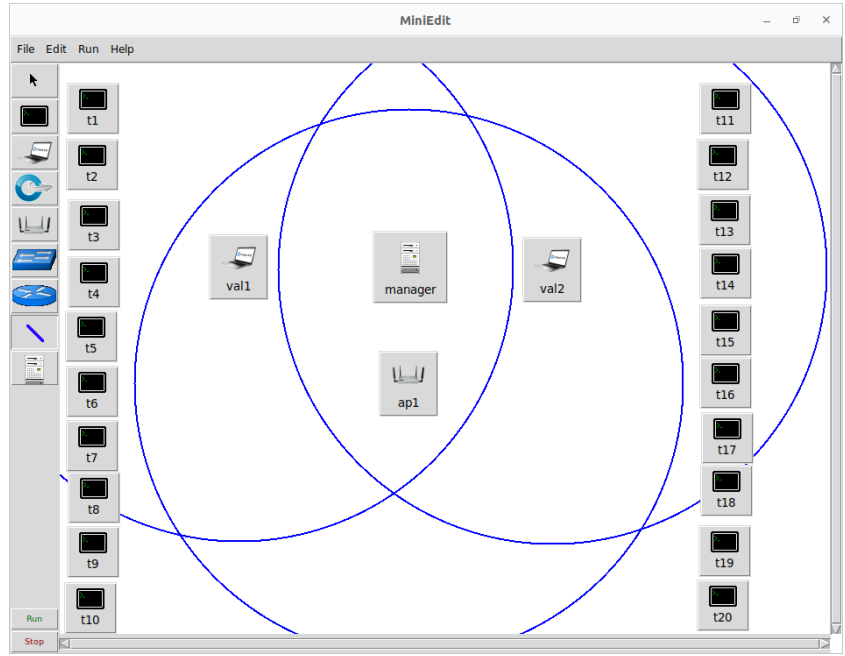

**Fig. 5.** Experiment Topology

For the evaluation of our proposal, we use Mininet and its wireless extension Mininet-WiFi as the emulator, which can be used to emulate Software Defined Wireless Networks. Mininet-WiFi enhances Mininet emulator with virtual wireless stations and access points [18]. The emulator also allows the execution of external code in the nodes, and through that, it is possible that we execute the most varied applications within the simulation.

We have built a topology of mobile nodes to simulate edge devices working as trainers in the system. In our initial topology, we created twenty trainer nodes and two validator nodes besides the global model manager. The twenty nodes are split evenly across two groups and each validator oversees a separate group. The validators are directly connected to the global model manager. As we do not have the rewards system working yet, the global model manager is who miners the new blocks in the blockchain for the moment. Figure 5 shows the topology created.

For blockchain interaction in a scalable way we have used Cartesi Rollups [3]. Layer-2 Cartesi can communicate with layer-1 either being the Ethereum mainet or a testnet. In our case we are using the Goerli testnet [26]. Goerli testnet has migrated from PoA to PoS as consensus mechanism to to mirror the main network and is the default and recommended network for executing testing of staking and validating. Our application already records the aggregated weights for each round in a file and we developed a dapp using Cartesi Machine that write the following data in the blockchain: *timestamp, stake of each miner, stake of each validator, Accumulated accuracy of each round* [20], avoiding scalability and storage space problems that would be common problems in using a public blockchain like Ethereum. As said before, Cartesi Rollups allows us to create our application in python and through the Cartesi Network we can execute transactions using the smart contracts already deployed on the blockchain.

Each client node in the simulation is programmed to execute the FL client code, sending its gradient to its specific validator. Each validator validates the model and if the results differ less that 25%, the result is marked as Positive. After calculate the global model using the Positive contributions sent by the trainers, the validators calculate the final model using the average of the models of the two validators, after that the result is recorded in the blockchain created in the simulator.

When a malicious device becomes a miner, it may try to disrupt the global model's calculation by putting bogus voting results on the system. These byzantine faults are captured by adding Gaussian noise to each trainer local model updates with probability 0.05 [22].

The next stage consists of the complete implementation of the reward mechanisms and the analysis of the improvement of accuracy in relation to an implementation without these mechanisms as well as the integration between these mechanisms and the reward mechanism used by Cartesi for the Fraud Proof [42].

## 6 Final Considerations

In this paper we present an intelligent framework for compliance verification of QoS metrics agreed via SLA by using blockchain technology to provide reliable transactions in a FL system. FL enables the ML training without exposing its valuable data to the public. while the recorded data cannot be altered because the data immutability in the blockchain. Blockchain is also be used as an incentive platform to encourage trainers to contribute honestly to improve and maintain the reliability and sustainability of the system using a optimistic rollup solution named Cartesi Rollups. In a nutshell, our contributions to increase the system security take place on two fronts: how to access data off-chain in a reliably way and how to prevent malicious trainers from harming the correct calculation of the ML model, that will be our future work in the next phase of the project.